\documentclass[fleqn,usenatbib,useAMS]{mnras}

\usepackage{graphicx}	
\usepackage{amsmath}	
\usepackage{amssymb}	
\usepackage{multicol}   
\usepackage{bm}		    
\usepackage{pdflscape}	
\usepackage{xcolor}	    
\usepackage{soul}

\usepackage[T1]{fontenc}
\usepackage{ae,aecompl}

\usepackage{newtxtext,newtxmath}

\title[White Dwarf Pollution: One Star or Two?]{White Dwarf Pollution: One Star or Two?}

\author[H.~T.~Noor et al.]{
Hiba Tu Noor,$^{1}$\thanks{E-mail: hiba.noor.19@ucl.ac.uk} Jay Farihi,$^{1}$ Mark Hollands,$^{2}$
and Silvia Toonen$^{3}$
\\
$^{1}$Department of Physics and Astronomy, University College London, London WC1E 6BT, UK\\
$^{2}$Department of Physics and Astronomy, University of Sheffield, Sheffield, S3 7RH, UK\\
$^{3}$Anton Pannekoek Institute for Astronomy, University of Amsterdam, 1090 GE Amsterdam, The Netherlands}

\date{Accepted XXX.  Received YYY; in original form ZZZ}

\pubyear{2023}

\begin{document}
\label{firstpage}
\pagerange{\pageref{firstpage}--\pageref{lastpage}}
\maketitle

\begin{abstract}
The accretion of tidally disrupted planetary bodies is the current consensus model for the presence of photospheric metals commonly detected in white dwarfs.  While most dynamical studies have considered a single star and associated planetary instabilities, several investigations have instead considered the influence of widely-bound stellar companions as potential drivers of white dwarf pollution.  This study examines the prevalence of wide binaries among polluted white dwarfs using {\em Gaia} Data Release 3 astrometry, where three samples are investigated: 71 DAZ stars with metals detected in the ultraviolet using {\em Hubble
Space Telescope}, and two groups of DZ stars identified via Sloan Digital Sky Survey spectroscopy, comprised of 116 warmer and 101 cooler sources.  Each sample was searched for spatially-resolved, co-moving companions, and compared to the same analysis of thousands of field white dwarfs within overlapping regions of the {\em Gaia} Hertzsprung–Russell diagram.  The wide binary fraction of the DAZ sample is $10.6_{-3.2}^{+3.9}$\,per cent, and within $1\upsigma$ of the corresponding field.  However, the search yields wide binary fractions of less than 1.8\,per cent for the two independent DZ star catalogues, which are each distinct from their fields by more than $3\upsigma$.  Both sets of results support that pollution in white dwarfs is not the result of stellar companions, and the delivery of metals to white dwarf surfaces is caused by major planets.  The discrepancy between the DAZ and DZ star wide binary fractions cannot be caused by white dwarf spectral evolution, suggesting these two populations may have distinct planetary architectures.
\end{abstract}




\section{Introduction}
White dwarfs provide a unique opportunity to study exoplanetary composition and dynamics.  Studies of white dwarf spectra have found between one-quarter and one-half of typical and relatively cool white dwarfs exhibit atmospheric metals that should otherwise sink due to efficient gravitational settling \citep{Zuckerman_2010, Koester_2014}.  Given that the atmospheric diffusion timescales are always orders of magnitudes shorter than their evolutionary (cooling) ages, pollution in white dwarfs must have occurred in their recent histories and is often ongoing \citep{Paquette_1986, Koester_2009}.

The source of photospheric metals in white dwarfs has been decisively demonstrated to be consistent with the accretion of tidally-disrupted planetary bodies \citep{Jura_2003, Farihi_2009}.  Compelling support for this scenario comes through observations of circumstellar debris disks that are now confirmed to closely orbit dozens of polluted white dwarfs \citep[see][]{Farihi_2016}.  In the vast majority of cases, the accreted material resembles the rocky bodies of the inner solar system, i.e.\ depleted in volatiles and dominated by O, Si, Mg, and Fe (the same four elements that comprise of 94\,per cent of the bulk Earth; \citealt{JuraYoung_2014}).  Some water-rich objects have also been identified \citep[e.g.][]{Farihi_2013, Raddi_2015}, including one case of a potential Kuiper belt analogue \citep{Xu_2017}.  

Although it is clear that metal pollution in white dwarfs is of planetary origin, the underlying mechanism to dynamically excite planetary bodies onto star-grazing orbits remains uncertain.  However, several compelling models have been proposed to account for this phenomenon.  The onset of dynamical instability following post-main sequence mass loss can scatter planetary bodies to star-grazing orbits where they can be tidally disrupted and accreted by the white dwarf.  The presence of at least one major planet can facilitate this process through mean-motion resonances \citep{Bonsor_2011, FrewenHansen_2014, AntoniadouVeras_2016}, secular resonances \citep{Smallwood_2018} and planet-planet scattering \citep{Payne_2016, Payne_2017}.  Such mechanisms, however, generally rely on short-timescale instabilities, and therefore struggle to account for pollution in white dwarfs with cooling ages of several Gyr.

Some studies have suggested the influence of wide
binary companions may significantly contribute to the observed pollution in white dwarfs.  A few variations of this possibility have been studied, including the influence of Galactic tides, such that during periods of close periastron passage, a wide companion is able to induce late-time instabilities in the white dwarf planetary system \citep{BonsorVeras_2015}.  Lidov–Kozai oscillations induced by a distant stellar companion have also been suggested to excite planetesimals on previously near-circular orbits to arbitrarily high eccentricities \citep{Stephan_2017, PetrovichMunoz_2017}.  Post-main sequence mass loss can further exacerbate this phenomenon, potentially bringing planetesimals within the white dwarf Roche limit to be accreted \citep{HamersZwart_2016}.  The distinctive feature that separates these studies from most other dynamical models is their ability to account for pollution at cooling ages beyond 1\,Gyr.  Even so, their contribution to white dwarf pollution is strongly dependent on the actual frequency of wide binaries orbiting white dwarfs.  

Results to date have found the frequency of atmospheric pollution among single white dwarfs to be indistinct from those in binary systems \citep{Wilson_2019, Zuckerman_2014} - findings that should be taken with caution as the sample statistics (12 and 38 binaries examined for pollution, respectively) are far from robust.  A search for infrared excesses around an unbiased sample of 14 polluted white dwarfs in wide binaries, revealed none to have a significant flux or colour excess, while an independent assessment using \textit{Gaia} Data Release 2 astrometry further confirmed a duplicity fraction consistent with zero for the 40 known dusty white dwarfs with infrared excesses \citep{Wilson_2019}.  Previous efforts to identify binary companions or duplicity fractions among white dwarfs have been extremely limited by incomplete or inhomogeneous proper motion and photometric surveys \citep{Farihi_2005,Hollands_2018_20pc}.  Far superior constraints on the wide binary population of polluted white dwarfs can now be obtained with \textit{Gaia}.  

In this work, \textit{Gaia} Data Release 3 (DR3; \citealt{GaiaDR3}) is utilised to elucidate the role of wide binaries in white dwarf pollution.  In the case that wide binaries are the major driver of the observed pollution, their frequency around polluted white dwarfs should be significantly greater than their occurrence around a random sample of white dwarfs.  This hypothesis is tested for three independently assembled catalogues of polluted white dwarfs.  The search is conducted to wide separations, such that at the largest of separations, the binary catalogue is dominated by chance alignments.  Furthermore, bound probabilities are assigned to each candidate binary, thus making it possible to determine high-confidence binary fractions of each sample.

The remainder of this work is organised as follows.  Section~\ref{sec:sample selection} describes the target white dwarf and field selections as well as the strategy for minimising contamination from chance alignments.  Section~\ref{sec:wide binary criteria} details the wide binary search technique and how the bound probabilities are quantified for candidate binaries through the use of Gaussian statistics.  Key findings are outlined and their implications discussed in Section~\ref{sec:binaryfractions}.  A summary is presented in Section~\ref{sec:conclusion}.

\section{Sample Selection and Processing}
\label{sec:sample selection} 

Below are described the selected samples of polluted white dwarfs, their corresponding \textit{Gaia} assessment, as well as the chosen method of selecting field stars whose wide binary fraction can be compared to that of the polluted white dwarf samples.

There are three samples of metal-rich white dwarfs utilized to study the fraction of wide binary companions that may dynamically influence their respective planetary systems and pollution characteristics.  The first sample is a subset of stars with photospheric metal detections listed in table~A1 of \citet[hereafter W19]{Wilson_2019}.  This list contains a total of 154 DA-type\footnote{Spectral type DA denotes white dwarfs whose strongest spectral features are the Balmer lines, that typically signify a hydrogen-rich atmosphere, where DAZ denotes weaker lines of metals are present.  In contrast, the DZ spectral class are stars with only metal absorption features, where typically the atmosphere is helium dominated.} white dwarfs observed in the ultraviolet, 71\footnote{Excluding the magnetic object WD\,1533--057, as it may be a merger product, and the Si\,{\sc ii} lines are likely interstellar (B. T.~G\"ansicke 2023, private communication).} of which are reported to exhibit photospheric lines of Si\,{\sc ii} and are thus DAZ stars. These stars have 14\,000\,K $< T_{\rm eff} < 31\,000$\,K, and cooling ages in the range 9--300\,Myr.  

The remaining two samples are cooler DZ stars identified using the Sloan Digital Sky Survey (SDSS), typically by their strong Ca\,{\sc ii} H \& K absorption features. The first is a relatively warm (6000\,K $< T_{\rm eff} < 12\,000$\,K) sample of 146 white dwarfs from \citet[hereafter D07]{Dufour_2007}, and the second is a relatively cool (4500\,K $< T_{\rm eff} < 9000$\,K) group of 231 stars from \citet[hereafter H18]{Hollands_2018_DZ}, with four stars in common between the two samples.  These two samples have some modest overlap in their evolutionary ages, which span the range 400\,Myr -- 8\,Gyr.

For the purpose of this study, all sources are required to have the 5-parameter astrometric solutions in \textit{Gaia} DR3, i.e.\ a successful measurement of the its 2-D position on the sky, parallax, and 2-D proper motion.  To minimise contamination from chance alignments, the binary search is restricted to stars with total proper motion, $\upmu>5$\,mas\,yr$^{-1}$.  Both members of a candidate binary are further required to have relatively precise astrometry such that the fractional error in parallax, $\varpi/\upsigma_{\varpi}> 10$.  Any star that does not satisfy the preceding quality cuts is removed, reducing the sample sizes to 71, 116, and 101 metal-rich stars in W19, D07, and H18, respectively.

Note that the quality cuts of the present study differ somewhat from previous works that utilise \textit{Gaia} astrometric searches for bound companions.  The practical effect of the cuts imposed on the purity and completeness of the samples will thus also be modestly different.  Typical examples include a more stringent cut of $\varpi/\upsigma_{\varpi}> 20$ for the primary star, but more liberal cuts of $\varpi/\upsigma_{\varpi}> 5$ (or even 2) for the companion \citep{ER18,Tian_2020}.  The first condition would degrade the statistical confidence of determined binary fractions by significantly reducing the number of sources studied, while the second condition would likely increase chance alignments.  Based on trial and error, a more stringent cut on candidate companion parallax was found to exclude a modest fraction of genuine binaries, while reducing the number of chance alignments.  At least one study has insisted on stars that have $\upmu>40$\,mas\,yr$^{-1}$ \citep{HartmanLepine_2020}, but again, this would negatively impact sample sizes in the present study.

\begin{figure*}
\includegraphics[width=\textwidth]{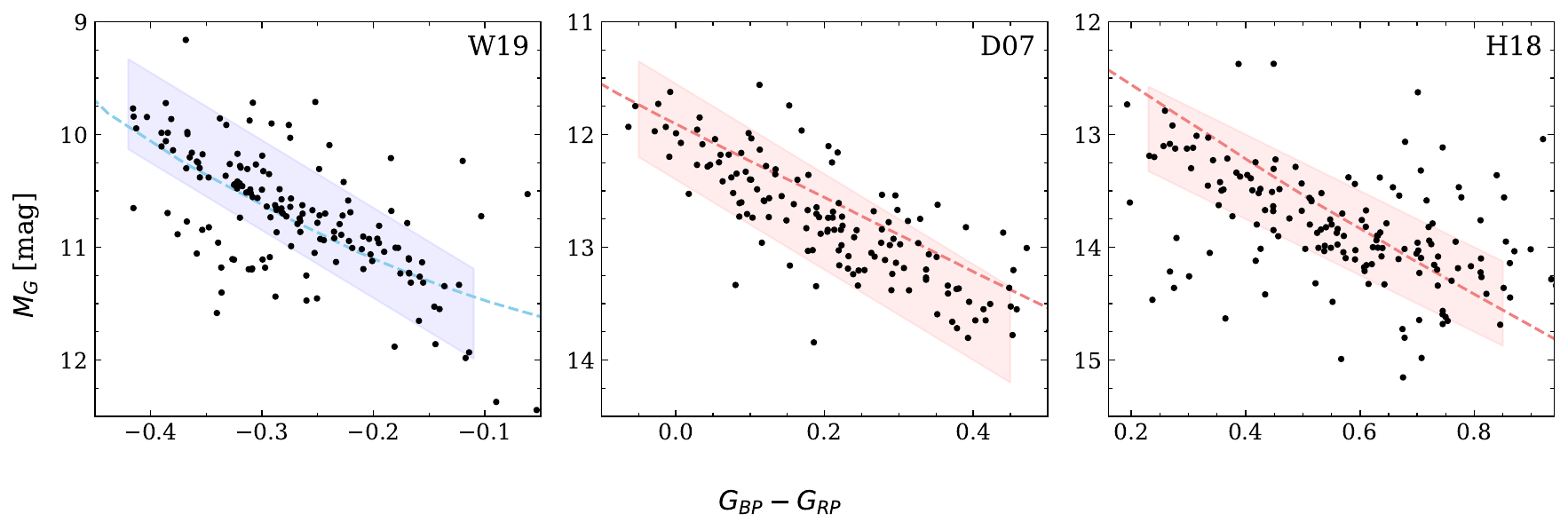}
\vskip 0 pt
\caption{White dwarfs in the W19, D07, and H18 samples, plotted on a colour-absolute magnitude diagram using \textit{Gaia} photometry.  The blue and red dashed lines represent white dwarf cooling sequences for pure hydrogen and helium atmospheres, for stellar mass 0.6\,M$_{\odot}$.  The shaded parallelograms were defined to select field white dwarfs, where any \textit{Gaia} sources within the enclosed region can form part of the appropriate three fields.}
\label{fig:fields}
\end{figure*}

\textit{Gaia} DR3 is further used to construct three groups of field stars whose wide binary fractions can be compared to that of each polluted white dwarf sample.  To choose white dwarfs in the field, a parallelogram is constructed around each science target sample locus within the colour-absolute magnitude diagram (i.e.\ similar $G_{\rm BP} - G_{\rm RP}$ colour and $G$ magnitude), as illustrated in Figure~\ref{fig:fields}.  A relatively simple query of the DR3 database can be made with these constraints, as can be found in the Appendix.  These queries return 15\,248 field white dwarfs for the W19 sample (Field 1), 31\,421 objects for the D07 stars (Field 2), and 28\,597 sources for the H18 targets (Field 3).  From the latter two, a random sample of 16\,000 white dwarfs is taken for two reasons; 1) query execution times during the subsequent binary search can exceed 10\,h for larger sample sizes, and 2) a random sample of 16\,000 will be statistically robust and representative. 

To ensure that the field samples do not contain any contaminants that may appear in this part of the HR diagram, the catalogue of white dwarf candidates by \citet{GentileFusillo_2021} is utilised. Any field star that is not included in this catalogue is removed, resulting in reduced sample sizes of 15\,218, 15\,917, and 15\,779 white dwarfs in Fields 1, 2, and 3, respectively.  The wide binary search is executed identically on each of the three target catalogues, and their corresponding fields.

This study requires a direct comparison of the polluted white dwarf wide binary fraction with that of similar field stars.  It is thus imperative to ensure that the target samples are at least as sensitive to the detection of wide binaries as their comparison fields.  This is essential to avoid under- or over-counting binary systems within the science or field samples, thereby preserving the integrity of the comparative analysis.  This sensitivity depends on the apparent $G$-band magnitude of the star and its line of sight distance, because brighter and more nearby stars have more precise astrometry than fainter and more distant stars.

\begin{figure*}
\includegraphics[width=\textwidth]{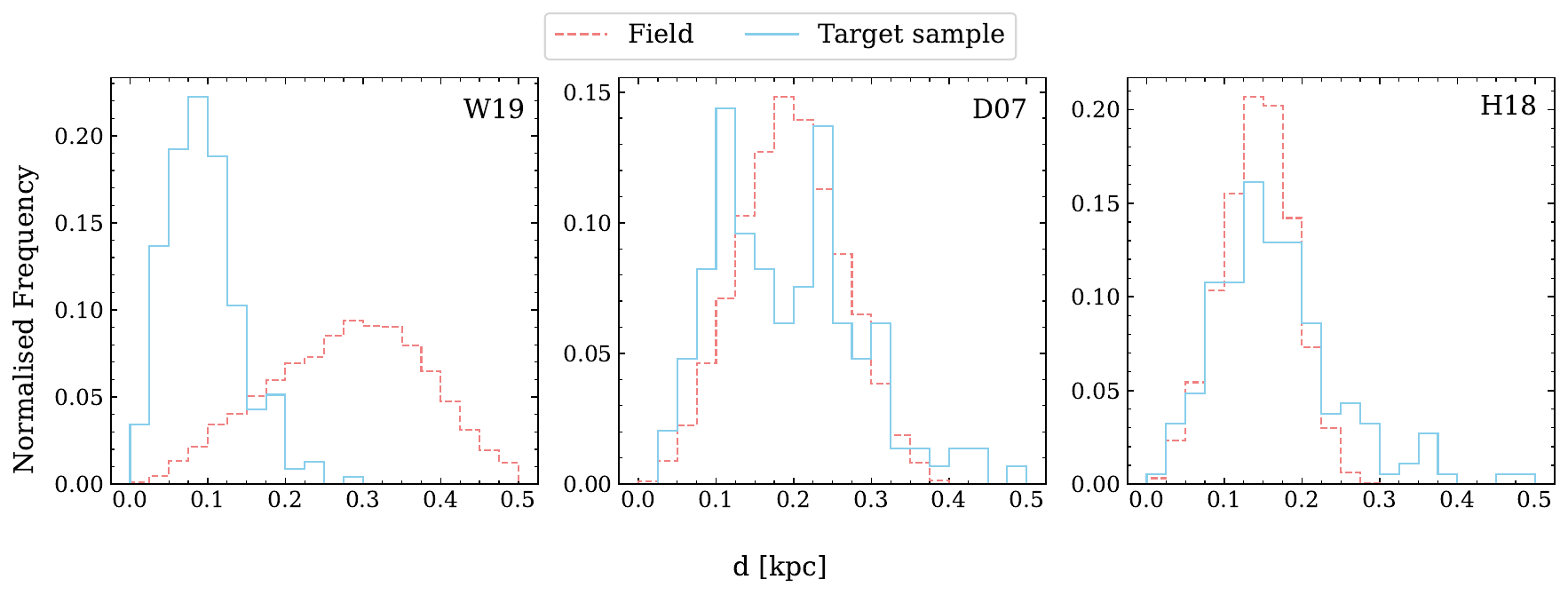}
\vskip 0 pt
\caption{Normalised histograms of line of sight distance for the target catalogue stars (blue) and their corresponding fields (red).  The D07 and H18 target samples probe similar distances to their fields, while the W19 group is nearer than its field counterparts, and is hence more sensitive to the detection of wide binaries.  }
\label{fig:fieldvtarget}
\end{figure*}

Figure~\ref{fig:fieldvtarget} illustrates that this requirement is satisfied.  The D07 and H18 science samples have distance distributions that are similar to those of their corresponding fields, and thus a compatible sensitivity to the coolest M dwarf and white dwarf companions.  And while the W19 target sample has a distribution of distances that does not follow its field star counterparts, the science targets are closer, and hence moderately more sensitive to the detection of wide binaries.  If anything, the field counterparts may be modestly under-counted.

\section{Wide binary search and analyses}
\label{sec:wide binary criteria}


\subsection{Results of astrometric criteria}
\label{sec:wide binary search}

For each white dwarf science target or field star, a candidate companion is required to lie within a projected separation radius $s = 1$\,pc (206\,265\,AU).  The search is restricted to this radius as the contamination rate from chance alignments is expected to grow rapidly beyond this cutoff, and a negligible number of bound binaries are expected to exist at wider separations, where the Galactic tidal field overwhelms the gravitational attraction between two stars \citep{BinneyTremaine_2008}.  

Further selection criteria include the requirement that a pair of stars have consistent parallaxes, and also proper motions consistent with a gravitationally bound orbit (equations~2 and 7 in \citealt{ER18}).  After selection, the candidate binary samples are cleaned from higher order multiples, ensuring that if either the white dwarf or its companion forms part of another candidate binary, both pairs are removed.  It is important to note some real binaries will be removed during this cleaning process, e.g.\ a tertiary candidate that is actually a chance alignment may cause a genuine binary to be rejected.  This was not an issue for the target samples, as no resolved triples were retrieved, or the fields, where the number of rejected candidate triples were less than 0.6\,per cent of the total sample.  Unavoidably, a small number of unresolved hierarchical triples are also expected to satisfy the wide binary criteria, and incorrectly be assigned as a binary system.  Contamination from these systems is discussed in Section~\ref{sec:chance alignments}.

The binary search returns 1333, 1465, and 1571 candidate binaries in Fields 1, 2, and 3.  For the science samples, the search returns 12 candidate binaries among the stars in W19, but no co-moving sources for the DZ white dwarfs in D07 and H18. Because of this unexpected result for the DZ stars, and as an independent check to  ensure no binaries were missed as a result of the astrometric criteria imposed, the companion search was re-run for all three polluted white dwarf samples.  In this ancillary search, the condition that any proper motion difference should be within $3\upsigma_{\upDelta\upmu}$ of a bound orbit, was relaxed to $10\upsigma_{\upDelta\upmu}$.  Note that this reduces the possibility of missing true binaries in the science samples, but increases the number of chance alignments.  This weaker criterion returns no additional co-moving sources in W19, but recovers five and three candidate binaries in D07 and H18, respectively.  These candidates are discussed in Section~\ref{sec:binaryfractions}, but are strictly supplemental to the main results.

\subsection{Limitations}

It is important to note limitations that may affect the binary search.  The criteria assume the semi-major axis of the orbit is equal to the projected separation of the binary, and this assumption may not hold, but will be correct within a factor of two in most cases \citep[for a detailed discussion, see][]{ER18}.  

Furthermore, the difference in tangential velocities is assumed to be equal to the physical velocity difference between the two stars as resulting from orbital motion by Kepler's third law.  However, at larger angular separations, projection effects become increasingly important, and this assumption will eventually break down \citep{ElBadry_2019}.  While this can be corrected if the radial velocity of at least one binary component is known, such measurements are not available for most candidate binaries in this study.  On the other end of angular separations, the spatial resolution of {\em Gaia} prevents the detection of close binaries, where the effective limit is larger at greater distances from the Sun.  Lastly, the {\em Gaia} magnitude limit of $G\approx20.7$\,mag prevents the detection of the faintest companions.  Nevertheless, these limitations are expected to affect the science target and field samples similarly, as Figure~\ref{fig:fieldvtarget} nicely demonstrates, so that a comparison between the two should suffice to test the hypothesis that is the subject of this work.

\subsection{Chance alignment probabilities}
\label{sec:chance alignments} 

Most candidate binaries that satisfy the initial criteria are expected to be true bound sources.  However, given the large search radius, the inclusion of chance alignments is almost certainly guaranteed.  A robust methodology is thus required to distinguish the two populations.

Although it is difficult to ascertain whether a given candidate binary is a chance alignment, it is possible to constrain contamination from chance alignments in a probabilistic manner.  To estimate the probability of a candidate pair being a chance alignment, the methodology of \citet{Lepine_2007} is followed, where chance alignment catalogues are constructed for each field.  The position of each field white dwarf is artificially shifted by $1\degr$ from its actual location on the sky.  The binary search is then repeated with these artificial coordinates, and any candidate companions that are retrieved as a result are catalogued as chance alignments (because the pair of stars are too distant to be bound).  However, this approach will not be fully valid for the widest and nearest binaries with angular separations $\theta>1\degr$ and distances $d<60$\,pc.  Because such pairs may continue to appear in the chance alignment catalogue even when the celestial coordinates of one binary component are shifted by $1\degr$, they are removed from the process.  Overall, this procedure successfully eliminates genuine binaries (since white dwarfs are shifted away from their true companions), while also preserving chance alignment statistics that are expected to vary little over $1\degr$ scales.  This results in chance alignment catalogues with 234, 286, and 279 pairs for Fields 1, 2, and 3, respectively.

\begin{figure}
\includegraphics[width=\columnwidth]{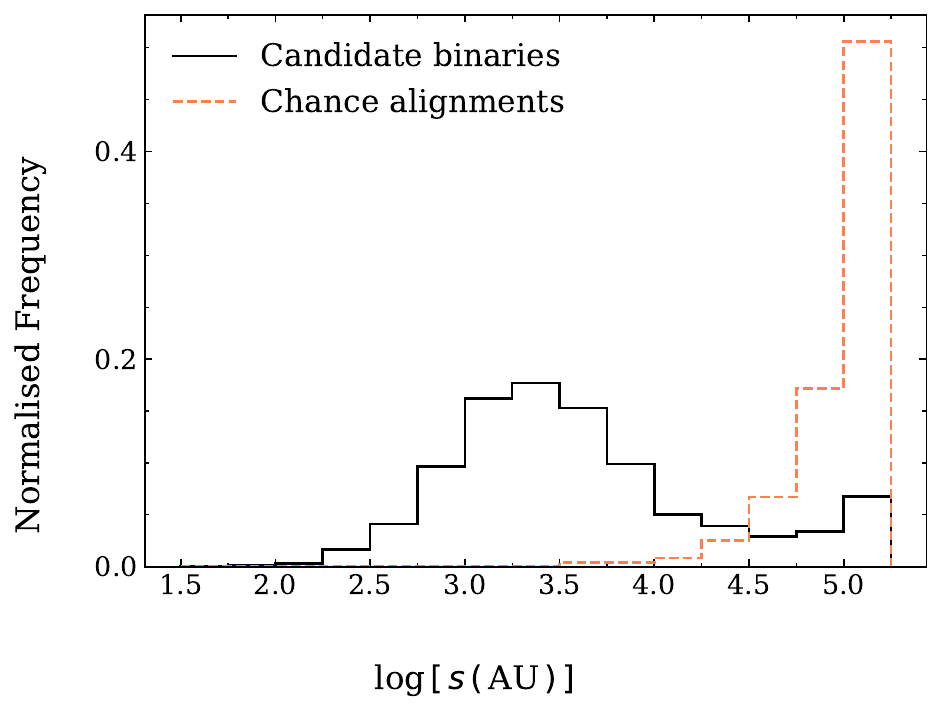}
\vskip 0pt
\caption{Normalised distribution of projected separation for candidate binaries and chance alignments in Field 1.  Candidate binaries have high purity at $s \lesssim 10^4$\,AU, but quickly become dominated by chance alignments at greater separations. Highly similar distributions result for Fields 2 and 3.}
\label{fig:chancevcandidates}
\end{figure}

Figure~\ref{fig:chancevcandidates} compares the separation distributions of candidate binaries in Field 1 to pairs in the corresponding chance alignment catalogue (containing no genuine binaries).  As expected, chance alignments dominate the candidate binaries at the widest separations ($s\gtrsim10^{4.5}$\,AU).  Note that while most candidate binaries in this regime are chance alignments, no cut is enforced on the allowed projected separation to avoid missing the widest authentic binaries.  Although chance alignment probabilities can be determined using a Bayesian formulation \citep{Andrews_2017}, the sample sizes in the present study are inadequate for the construction of a reliable inference model.

Following \citet{ERH_2021}, the probability of a given pair of stars being a chance alignment is estimated by comparing the local density of candidate binaries to that of chance alignments in relevant parameter spaces.  Five parameters are adopted that were chosen through trial and error, these are; the projected separation, the parallax difference and its uncertainty, the parallax uncertainty of the white dwarf, and the local sky density.  The latter can be calculated by counting the total number of sources per square degree around each white dwarf that (a) pass the quality cuts, and (b) have $G < 20$\,mag.  The density kernel widths are chosen to have $\upsigma = 0.3$, 0.35, and 0.35 for Fields 1, 2, and 3, such that these are (i) sufficiently wide to avoid over-fitting, and (ii) sufficiently narrow to avoid over-smoothing.  

For a candidate binary that takes location $\Vec{x}$ in the 5-dimensional parameter space, the local density of chance alignments is expressed as $\mathcal{N}_{\rm chance}(\Vec{x})$, and that of binary candidates as $\mathcal{N}_{\rm cand}(\Vec{x})$.  The chance alignment probability statistic for the candidate binary is then the ratio of these two, i.e.\ $\mathcal{R}(\Vec{x}) = \mathcal{N}_{\rm chance}(\Vec{x}) / \mathcal{N}_{\rm cand}(\Vec{x})$.  Owing to the modest number of candidate binaries retrieved in the target samples, it is not possible to determine $\mathcal{N}_{\rm cand}$ for these, and the corresponding field is used for this purpose.  By understanding that the distributions of target white dwarfs are similar to those of the field, the required densities for $\mathcal{R}(\Vec{x})$ are calculated by evaluating field distributions at the locations of the science targets in the 5-dimensional parameter space.

\begin{figure*}
\includegraphics[width=\textwidth]{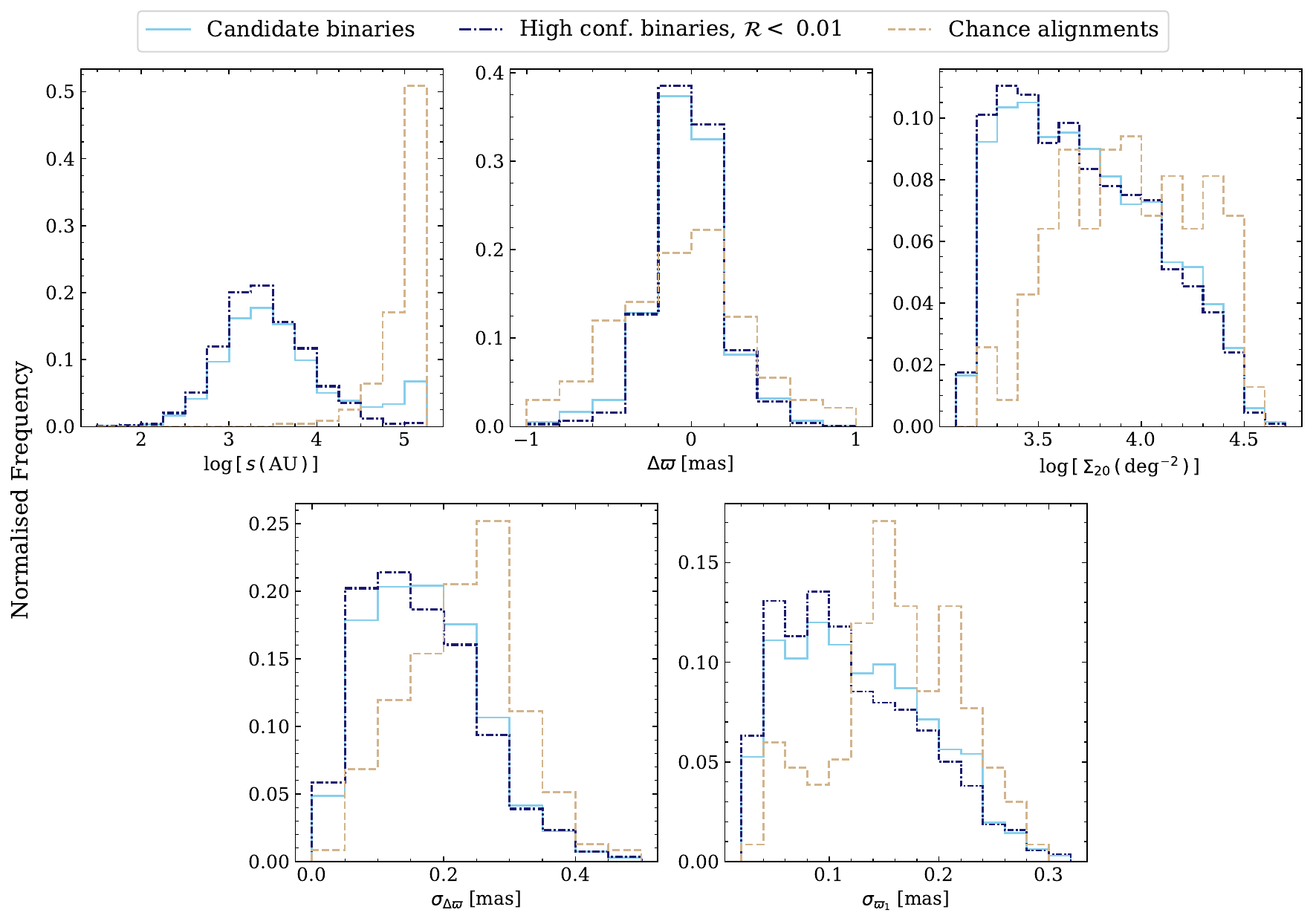}
\vskip 0pt
\caption{Normalised distributions of chance alignments, candidate binaries, and high-confidence binaries in Field 1 for the parameters defining the Gaussian KDEs.  In comparison to the high-confidence binaries, chance alignments are generally located in regions of higher local sky density, have greater projected separations, larger differences in parallaxes and higher uncertainties. Highly similar trends result for Fields 2 and 3.}
\label{fig:distributions}
\end{figure*}

This method does not guarantee the inclusion of every genuine binary, as the distributions of the chosen parameters are not strictly Gaussian (Figure~\ref{fig:distributions}).  Furthermore, the estimated densities from the KDE, and thus the chance alignment probabilities, are strongly dependent on the choice of bandwidth, more so than the kernel shape.  Nevertheless, while great care has been taken to select the best parameters and bandwidth, this study makes no claim that these are optimal for general usage.

Note that $\mathcal{R}$ is not strictly a probability, as its value can exceed one.  Although candidates with $\mathcal{R} \gtrsim 1$ are likely chance alignments and can be excluded from further consideration, in the case of $\mathcal{R} <1$, the statistic can be interpreted as a tracer for the actual chance alignment probability.  High-confidence binaries are defined here as those with $\mathcal{R} < 0.01$, corresponding to where one expects better than 99\,per cent probability of being bound.  This threshold is chosen as a trade off between minimising contamination from chance alignments, and maximising the number of genuine binaries retrieved across the samples.

\begin{figure}
\includegraphics[width=\columnwidth]{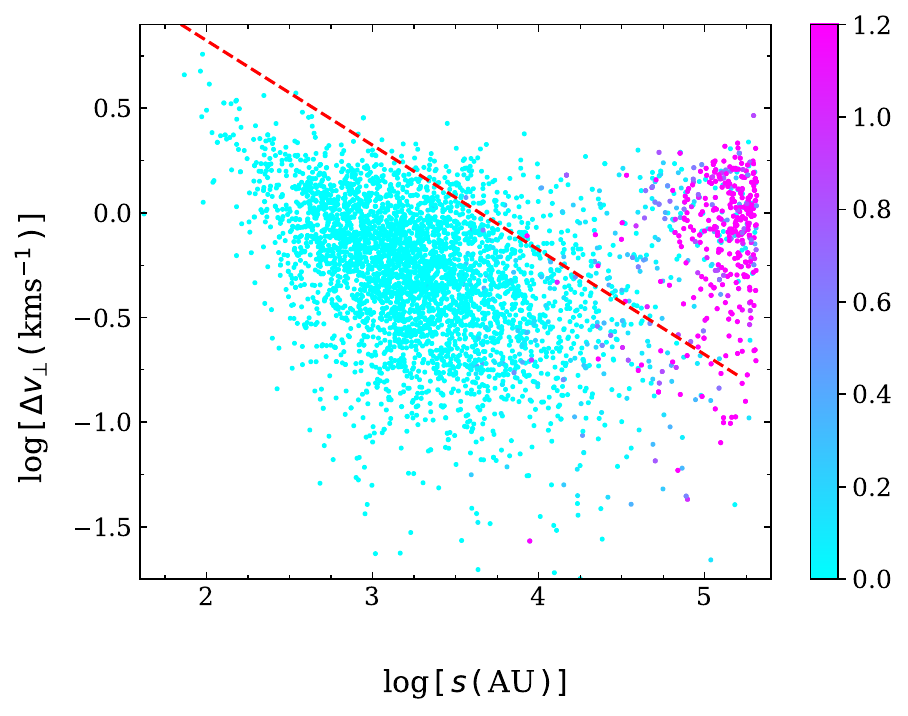}
\vskip 0pt
\caption{Tangential velocity difference as a function of projected separation for all candidate binaries in this work coloured by the magnitude of $\mathcal{R}$.  The red dashed line represents the maximum tangential velocity difference, as predicted by Kepler's third law for a circular orbit of total system mass 5\,M$_{\odot}$.  Most pairs at large separation are chance alignments, and are thus excluded by the requirement of $\mathcal{R}<0.01$.}
\label{fig:highconfbinaries}
\end{figure}

Figure~\ref{fig:highconfbinaries} explores tangential velocity difference as a function of projected separation for all binary candidates, and the effect of the $\mathcal{R}$ cut on the binary sample purity and completeness.  As expected, the vast majority of candidate binaries at close separations have low $\mathcal{R}$ values, and hence high probabilities of being bound.  At the largest separations, most candidates are chance alignments, with correspondingly high $\mathcal{R}$ values.  These are primarily pairs with large proper motion uncertainties, otherwise they would be excluded by the initial conditions imposed in the binary search.  Moreover, there exists a population of high-confidence binaries that have tangential velocity differences greater than those predicted by Kepler's third law; these could be a result of higher proper motion uncertainties, projection effects, or otherwise non-circular orbits \citep{ERH_2021}.  This population may also consist of triples, where the tangential velocity of one component can be affected by the gravitational effects of an unresolved companion \citep{Belokurov_2020}.  Studies that prioritise sample purity over completeness should consider removing these systems, because their ambiguous classification as unresolved triples may affect the accuracy of subsequent analyses.

\section{Wide Binary Fractions and Discussion}
\label{sec:binaryfractions}

Table~\ref{tab:binary_fractions} lists the results of this study, the high-confidence binary fractions found for all three target samples of polluted white dwarfs, and their corresponding field stars.  Binary fractions and their respective uncertainties are determined by using a $\upbeta$ probability distribution corresponding to the binomial likelihood of the data with a flat prior.  The $16^{\rm th}$ and $84^{\rm th}$ percentiles are adopted as the $1\upsigma$ limits, and the binary fraction itself is taken as the $50^{\rm th}$ percentile value.  For samples with no binaries, the $84^{\rm th}$ percentile is used as a $1\upsigma$ upper limit for consistency.  For the W19 sample, seven of 71 polluted white dwarfs are in wide binary systems, and five of 82 stars that are not polluted have wide companions.  Both of these fractions are consistent with each other, as well as with the corresponding field wide binary fraction within $1\upsigma$.  Thus, it appears that wide binaries do not occur more frequently around these DAZ white dwarfs than they occur for the DA sample or field white dwarfs.  

The three control samples of field white dwarfs have wide binary fractions that are broadly in agreement; there is less than $3\upsigma$ difference between the values for Fields 1 and 2, and similarly for Fields 2 and 3, where the fractions for Fields 1 and 3 are within $5\upsigma$.  These field binary fractions modestly increase with decreasing sample distance (Figure~\ref{fig:fieldvtarget}; $d_1=0.29\pm0.10$\,kpc, $d_2=0.20\pm0.07$\,kpc, $d_3=0.14\pm0.05$\,kpc), suggesting a modest luminosity bias is present, and thus the Field 1 and 2 wide binary fractions are moderately underestimated.

An estimate of missed binaries in the more distant Fields 1 and 2, relative to Field 3, is derived as follows.  The differences between their median line of sight distances are computed, and constants corresponding to these distance moduli are added to the apparent $G$ magnitudes of the Field 3 white dwarfs.  In cases where the new $G$-band magnitudes are fainter than 20.6\,mag (the faintest companions detected in all three fields), they are now classified as missed detections.  The results are that 294 and 86 of the 1322 high-confidence, wide binaries in Field 3 would have companions below the detection threshold if they were as distant as Fields 1 and 2, respectively.  That is, only 78\,per cent and 93\,per cent of the Field 3 companions would be detected at the distances to Fields 1 and 2, respectively.

A commensurate number of companions have thus almost certainly been missed in the more distant Fields 1 and 2.  Inverting the fractions obtained above, more accurate wide binary fractions for Fields 1 and 2 are 9.10\,per cent and 8.06\,per cent, respectively.  All three field binary fractions are in agreement within $3\upsigma$, and thus Field 3 is almost certainly the most representative of the actual wide binary fraction of field white dwarfs.

\begin{table}
\begin{center}	 
\caption{High-confidence, wide binary fractions for all samples.}
\label{tab:binary_fractions}
\begin{tabular}{llllr} 

\hline
Sample      &$N$        &$n$                                &SpT    &Wide binary fraction\\
            &           &                                   &       &(per cent)\\
\hline

W19         &71         &7                                  &DAZ    &$10.6(-3.2,+3.9)$\\
W19         &82         &5                                  &DA     &$6.8(-2.4,+3.1)$\\

Field 1     &$15\,218$  &1078                               &       &$7.08(-0.20,+0.21)$\\
Field 1     &\multicolumn{2}{l}{{\em --distance adjusted--}}&       &$9.10(-0.20,+0.21)$\\

\hline 
D07         &116        &0                                  &DZ     &$<1.6$\phantom{ )}\\
Field 2     &$15\,917$  &1200                               &       &$7.54(-0.20,+0.21)$\\
Field 2     &\multicolumn{2}{l}{{\em --distance adjusted--}}&       &$8.06(-0.20,+0.21)$\\

\hline

H18         &101        &0$^*$                              &DZ     &$<1.8$\phantom{ )}\\
Field 3     &$15\,779$  &1322                               &       &$8.38(-0.21,+0.22)$\\

\hline
\end{tabular}
\end{center}
{{\em Notes}.  $^*$See Section~\ref{sec:binaryfractions} for one exceptional case.}
\end{table}

The results for the two optical samples of DZ white dwarfs show a marked contrast to the ultraviolet sample of DAZ stars.  For both the D07 and H18 samples, the standard criteria searches yield no candidate binary companions, each inconsistent with those of their corresponding fields by at least $3\upsigma$.  This makes clear that wide companions do not contribute, even weakly, to the observed pollution in these DZ white dwarfs.  Given that no sample in this study exhibits a higher fraction of wide binaries for polluted white dwarfs, this finding suggests the delivery of material which pollutes white dwarfs is almost certainly driven by (as yet unseen) major planets.  

There are a handful of companion candidates that are worth discussing in more detail.  When the search criteria was eased for the DZ stars, so that their proper motion differences were allowed to be within $10\upsigma_{\upDelta\upmu}$ of a Keplerian orbit, the five candidate binaries thus retrieved for D07 all have $\mathcal{R}>1$, and are thus probable chance alignments.  Similarly, two of the three candidate binaries in H18 that resulted from the loosened criteria also have $\mathcal{R}>1$.  Only one DZ companion candidate has $\mathcal{R}<0.01$, and would otherwise be classified as a high-confidence binary:  SDSS\,J091621.36+254028.4 with companion LSPM\,J0916+2539 (\textit{Gaia}\,DR3\,687914402017314816).  The co-moving companion has an astrometric excess noise of 4.6\,mas, a RUWE of 4.8, and is thus an unresolved binary, making the system a hierarchical triple.  Nevertheless, it might be argued that the H18 DZ sample has a multiplicity fraction of $1.64_{-0.94}^{+1.50}$\,per cent, which is still drastically lower than for the field, or the DAZ stars. It is noteworthy that the wide companion of the DZ white dwarf SDSS\,J080740.68+493059.3 is not retrieved in this study, as the primary fails the quality cuts imposed in Section 2.

In \textit{Gaia} DR3, there exist intermediate separation binaries that are not spatially resolved but which can perturb the photocenter of the target star, and thus impact the astrometric solution.  Such potential companions would be at separations (e.g.\ within the orbit of Jupiter) that would dynamically preclude the formation of an S-type planetary system, but may still be interesting from the perspective of circumbinary (P-type) planetary system survival and pollution.  Thus, these candidate companions are not those widely-bound and postulated to gravitationally assist in white dwarf pollution as previously discussed.  Nevertheless, for completeness, the {\em Gaia} RUWE values of the target samples were examined, where a value near 1.0 indicates a good, single star solution and values above 1.4\footnote{ This threshold reduces to RUWE > 1.25 when considering DR3 data \citep{Penoyre_2022}, but has no effect on the findings here. } typically signify a binary source \citep{Belokurov_2020,Kervalla_2022}. There are no DZ white dwarfs with RUWE values that diverge 
significantly from unity (see the previous paragraph for a discussion of LSPM 
J0916+2539). Interestingly, there are three DAZ white dwarfs in W19 that have high RUWE values; WD\,0920+363 (1.74), WD\,1129+156 (6.49), and WD\,2231--267 (2.61), the latter of which is also in the non-single star catalogue of DR3.

While both sets of results for DAZ and DZ stars support that pollution in white dwarfs is not primarily driven by wide companions, the discrepancy between their wide binary fractions is difficult to understand {\em a priori}.  Apart from distinct $T_{\rm eff}$ and thus cooling ages (see Section~\ref{sec:sample selection}), the only clear difference between DAZ and DZ white dwarfs is their dominant atmospheric constituents; though not universal, DAZ stars typically have hydrogen-rich atmospheres, and DZ star atmospheres are usually helium-rich.  However, it does not appear possible to interpret their inconsistent duplicity on the basis of atmospheric composition alone.  Especially in light of the fact that it may be possible for white dwarf atmospheres to transition from hydrogen- to helium-rich, owing to the onset of convective mixing (depending on the initial hydrogen content, \citealt{Bedard_2023}).  Thus, theory may predict a common, or at least overlapping, origin for these two stellar populations.

The findings of this study are inconsistent with this picture at a fundamental level, at least insofar as these three metal-rich samples are representative of white dwarf spectral evolution.  The DZ stars in this study clearly do not have wide companions as commonly as do their field counterparts, or as do the DAZ stars studied here.  More specifically, these DAZ white dwarfs represent stars that do not later transform into DZ stars, and are thus inconsistent with spectral evolution changing their atmospheres from hydrogen into helium as they cool.

Stellar populations that are dominated by binaries are known, but there is no stellar spectral class previously known to be dominated by stars that evolved as singletons.  The detected common-proper motion pairs have projected separations $s>50$\,AU, and thus classical binary interactions (i.e.\ stable mass transfer, common-envelope) cannot have affected the white dwarf spectra nor their evolution.  However, singletons can in principle be the remnants of stellar mergers, where, theoretically, 10 to 30\,per cent of single white dwarfs are expected to descend from a merger \citep{Toonen_2017}, and that fraction may be higher among white dwarfs with relatively high masses \citep{Temmink_2020,Kilic_2023}.  Notably, there are indications that merger remnants may reside frequently in wide binaries \citep{Heintz_2022}, a possibility that is supported by synthetic models demonstrating that triples interact more often than binaries \citep{Toonen_2020,Shariat_2023}.  However, the possibility that DZ white dwarfs might be mergers is mentioned here only for completeness, and is otherwise severely problematic for myriad reasons, including but not limited to; mass distributions, kinematics, and in particular the orbits and detailed chemistry of the accreted planetary debris \citep{Koester_2014,Farihi_2016,Doyle_2019}.

The results here do not contradict the fact that a modest number of polluted white dwarfs are known to be in wide binary systems \citep{Zuckerman_2014}.  Previous studies have not performed blind searches as has been done here, and this is the first approach using sizable samples that are well-characterized and relatively homogeneous.  It is noteworthy that all known polluted white dwarfs with established infrared excesses are single stars, with the exception of the circumbinary dust disk orbiting the 2.3\,h binary SDSS\,J155720.77+091624.6 \citep{Farihi_2017,Wilson_2019}; none of these systems have any wide stellar companions that might be responsible for the atmospheric pollution.

A testable hypothesis that can account for the distinct wide binary fractions of the DAZ and DZ white dwarfs studied here, is the existence of separate sub-populations of polluted white dwarfs.  That is, a strongly polluted white dwarf population that is dominated by single stars, such as the optically-identified DZ stars and known, dusty stars with infrared excesses.  And then a second population that have wide companions as frequently as field white dwarfs, but with weaker metal pollution that requires sensitive detection in the ultraviolet with {\em Hubble}, such as the DAZ stars studied here.  Currently, there are insufficient unbiased samples of metal-enriched white dwarfs to test this possibility, but this may become feasible in the near future using the results of large spectroscopic surveys.  Such work might better constrain the efficiency of metal-enrichment among single white dwarfs compared to those that are in wide binaries.

\section{Conclusions}
\label{sec:conclusion}

This study aims to elucidate the role of wide binaries in white dwarf pollution, and utilises \textit{Gaia} DR3 astrometry to search for spatially-resolved, co-moving companions around three target and three control samples of white dwarfs.  For a relatively warm sample of DAZ white dwarfs identified by ultraviolet spectroscopy, the wide binary fraction is found to be roughly 10\,per cent and consistent with that of corresponding field stars with similar positions in the {\em Gaia} colour-absolute magnitude diagram.  For two cooler samples of DZ white dwarfs identified via SDSS spectroscopy, while their field star counterparts also show a wide binary fraction near 10\,per cent, there are no co-moving binary systems found by the search.  Neither of these results are consistent with models where wide stellar companions are the primary cause for white dwarf pollution, and instead the findings support major planets as the cause.

The resulting wide binary fraction of the DZ white dwarfs are distinct from field stars and also from the particular DAZ sample studied here.  This finding is consistent with a population dominated by stars that evolved as singletons, similar to that found so far for dusty white dwarfs with infrared excess.  It is suggested that polluted white dwarfs may belong to sub-groups; a highly polluted group dominated by single stars, and a modestly polluted group that has wide stellar companions as often as field white dwarfs.  Based on their inconsistent wide binary fractions, the DAZ stars studied here will not evolve into DZ stars through atmospheric mixing, placing some modest constraints on white dwarf spectral evolution models.

\section*{Acknowledgements}
The authors thank the anonymous referee, whose comments helped improve the manuscript. This work presents results from the European Space Agency space mission \textit{Gaia}, where data are processed by the \textit{Gaia} Data Processing and Analysis Consortium. MAH was supported by grant ST/V000853/1 from the Science and Technology Facilities Council (STFC). ST acknowledges support from the Netherlands Research Council NWO (VENI 639.041.645 and VIDI 203.061 grants).

\section*{Data Availability}
\textit{Gaia} data can be accessed through the \textit{Gaia} ESA Archive.  The high-confidence, wide binary and chance alignment catalogues for the control samples in Fields 1, 2, and 3 are available at \url{https://github.com/noorhiba/wdbinaries}. 



\bibliographystyle{mnras}
\bibliography{refs}




\appendix

\section{{\em Gaia} ADQL queries to assemble fields}
\label{sec:field query}

Fields 1, 2, and 3 can be assembled with the following query, \\
\texttt{SELECT * \\ FROM gaiadr3.gaia\_source \\
WHERE bp\_rp between A and B AND\\
phot\_g\_mean\_mag + (5 * log10(parallax) -10) < (C * bp\_rp) + D AND\\
phot\_g\_mean\_mag + (5 * log10(parallax) -10) > (E * bp\_rp) + F AND\\
parallax/parallax\_error > 10 AND\\
pm $>$ 5}\\
with \texttt{A},  \texttt{B},  \texttt{C},  \texttt{D},  \texttt{E}, and \texttt{F} constants as given in Table~\ref{tab:query constants}.

\begin{table}
\begin{center}	 
\caption{Constants required for the ADQL queries to assemble Fields 1, 2, and 3.}
\label{tab:query constants}
\begin{tabular}{ccccccc} 

\hline
Field   &\texttt{A}         &\texttt{B}         &\texttt{C} &\texttt{D} &\texttt{E} &\texttt{F}\\
\hline

1       &$-0.42\phantom{-}$ &$-0.11\phantom{-}$ &$6.00$     &$12.65$    &$6.00$     &$11.85$\\
2       &$-0.05\phantom{-}$ &$0.45$             &$4.00$     &$12.40$    &$4.00$     &$11.50$\\
3       &$0.23$             &$0.85$             &$2.50$     &$12.75$    &$2.50$     &$12.00$\\

\hline
\end{tabular}
\end{center}
\end{table}


\bsp	
\label{lastpage}
\end{document}